*Research Article*

# Estimating Finite Source Effects in Microlensing Events due to Free-Floating Planets with the Euclid Survey

**Lindita Hamolli,[1] Mimoza Hafizi,[1] Francesco De Paolis,[2] and Achille A. Nucita[2]**

[1]*Department of Physics, University of Tirana, Tirana, Albania*
[2]*Department of Mathematics and Physics "Ennio De Giorgi" and INFN, University of Salento, CP 193, 73100 Lecce, Italy*

Correspondence should be addressed to Lindita Hamolli; lindita.hamolli@fshn.edu.al

In recent years free-floating planets (FFPs) have drawn a great interest among astrophysicists. Gravitational microlensing is a unique and exclusive method for their investigation which may allow obtaining precious information about their mass and spatial distribution. The planned Euclid space-based observatory will be able to detect a substantial number of microlensing events caused by FFPs towards the Galactic bulge. Making use of a synthetic population algorithm, we investigate the possibility of detecting finite source effects in simulated microlensing events due to FFPs. We find a significant efficiency for finite source effect detection that turns out to be between 20% and 40% for a FFP power law mass function index in the range [0.9, 1.6]. For many of such events it will also be possible to measure the angular Einstein radius and therefore constrain the lens physical parameters. These kinds of observations will also offer a unique possibility to investigate the photosphere and atmosphere of Galactic bulge stars.

## 1. Introduction

Recent years have witnessed a rapid rise in the number of planetary objects discovered in the Milky Way with mass $M \leq 0.01 M_\odot$ that are not bound to a host star [1]. These objects are called free-floating planets (FFPs) or also rogue planets, nomads, or orphan planets (see [2] and references therein). Examples of objects of this kind are WISE 0855-0714, about 2 pc away from the Earth and Cha 110913-773444 (see http://exoplanet.eu). Due to their intrinsic faintness, it is very hard to detect them by direct imaging at distances larger than a few tens of parsecs. The only way to detect FFPs further away relies on the gravitational microlensing technique. This method has recently allowed the detection of ten FFPs towards the Galactic bulge, by selecting the microlensing events characterized by duration shorter than 2 days [3]. Indeed the event duration is proportional to the square root of the lens mass and therefore, based on statistical grounds, shorter events are caused by lower-mass lenses.

A consistent boost in the number of detectable FFPs is expected to occur by using space-based microlensing observations, in particular with the planned Euclid telescope.

Euclid is a Medium Class mission of the European Space Agency (ESA), which is scheduled to be launched in 2018-2019. Nowadays, the possibility to perform, by using the Euclid satellite, microlensing observations towards the Galactic bulge for about ~10 months [4] is under study. It is only marginally relevant, indeed, if the 10 months of observation is consecutive or not.

A gravitational microlensing event occurs when a massive object passes close enough to the line of sight to a distant source star and is described by three parameters: the time of maximum amplification $t_0$, Einstein time $T_E = R_E/v_T$ (where $R_E$ is the Einstein radius and $v_T$ is the transverse velocity between the lens and the source), and the impact parameter $u_0$ (the minimum value of the separation $u(t)$ between the lens and the line of sight in units of $R_E$). However, of these parameters only the Einstein radius crossing time, $T_E$, contains information about the lens and this gives rise to the so-called parameter degeneracy problem. One of the ways to break, at least partially, the microlensing parameter degeneracy is by considering the finite source effects induced in the microlensing light curves due to the finite extension of the source stars (that is when the source cannot be considered point-like, as in the Paczynski model) [5]. Finite source effects are nonnegligible when the value of $u_0$ becomes comparable to the source radius projected on the lens's plane in Einstein radius and the point-source approximation is not valid anymore [6]. These microlensing events are of particular importance due to various reasons. First, an event with a lens passing over a source star provides a rare chance to measure the brightness profile of a remote star. For such an event, different parts of the source star are magnified by different amounts. The resulting lensing light curve deviates from the standard form of a point-source event [7] and the analysis of the deviation may enable measuring the limb-darkening profile of the lensed star [8]. Second, in these kinds of events it is possible to measure the Einstein radius of the lens. The light curve at the moment of the entrance (exit) of the lens into (from) the source disk exhibits inflection of its curvature. The duration of the lens transit over the source $T$, as measured by the interval between the entrance and exit of the lens over the surface of the source star, is [9]

$$T = 2\sqrt{\rho_*^2 - u_0^2}\, T_E, \qquad (1)$$

where $\rho_*$ is the normalized source radius (in units of $\theta_E$). Once $T$ is measured and if $u_0$ and $T_E$ are known, the normalized source radius can be estimated through the relation in (1). With the additional information about the angular source size, $\theta_*$, the angular Einstein radius is measured as $\theta_E = \theta_*/\rho_*$. Since the Einstein radius does not depend on $v_T$, the physical parameters of the lens can be better constrained. Third, these events provide a chance to spectroscopically study remote Galactic bulge stars. Most stars in the Galactic bulge are too faint for spectroscopic observations even with large telescopes. However, the enhanced brightness of lensed stars of high-magnification events may allow performing spectroscopic observations, enabling population study of Galactic bulge stars [10]. These events might also allow performing polarimetric observations of selected ongoing events, for example, by using the VLT telescope, with the aim of further characterizing both the source and the lens system [11–13]. In the present paper we mainly concentrate on finite source effects on microlensing events caused by FFPs with the aim of obtaining a realistic treatment of the events, which are expected to be observable by the Euclid telescope. The great difference between space-based telescopes like the Euclid and usual ground-based microlensing observations is that the amplification threshold detectable by Euclid telescope is $A_{\rm th} = 1.001$ that, in turn, implies that the maximum value of $u$ turns out to be $u_{\max} = 6.54$ (much larger than the corresponding value for ground-based observation $u_{\max} = 1$).

In this respect, it is also important to emphasize that finite source effects are expected to occur and possibly be observable in a large number of microlensing events due to FFPs. Indeed, the smaller the lens mass is (therefore the smaller $\theta_E/\theta_*$), the more likely it is that these events involve source star disk crossing. The plan of the paper is as follows: in Section 2 we have shown the finite source microlensing equation considering different source limb-darkening profiles. Our main results are presented and discussed in Section 3, and we have summarized the main conclusions of this work in Section 4.

## 2. Finite Source Effects in FFP Microlensing Events

In order to make realistic treatment of the microlensing event rate due to FFPs expected to be observable by the Euclid telescope, it is important to consider finite source effects in the microlensing event light curves. Each light curve is calculated therefore by the intensity-weighted magnification averaged over the source disk [14]. For example, under the simplified assumption of a uniform source brightness and using polar coordinates centered at the source center, the magnification of an event by a finite source can be expressed as

$$A_{\rm fs}(u') = \frac{1}{\pi\rho_*^2} \int_0^{2\pi} d\theta \int_0^{\rho_*} r A(u')\, dr \qquad (2)$$

or, in extended form, by

$$A_{\rm fs}(u) = \frac{1}{\pi\rho_*^2} \int_0^{2\pi} d\theta \int_0^{\rho_*} r \\ \cdot \frac{u^2 + r^2 - 2ur\cos\theta + 2}{\sqrt{u^2 + r^2 - 2ur\cos\theta}\sqrt{u^2 + r^2 - 2ur\cos\theta + 4}}\, dr, \qquad (3)$$

where $\rho_* = R_* D_L / R_E D_S$ is the projected source radius in units of the Einstein radius $R_E$, $R_*$ is the physical source size, $u$ represents the normalized separation between the lens and the center of the source star, and $r$ (the radial coordinate in units of the Einstein radius) and $\theta$ are polar coordinates of a point on the source star surface with respect to the source star center. $D_S$ and $D_L$ are the source-observer and lens-observer distance, respectively. Figure 1(a) shows the magnification curve of a selected microlensing event when the source is assumed to be point-like (continuous line) and when it is considered extended, with a uniform brightness profile (dashed line calculated by (3)). Inside the Einstein radius, the residuals between the two curves may be significant. Therefore, by measuring the residuals with respect to the point-like case, one can estimate the angular Einstein radius of the microlensing event which may allow constraining the lens physical parameters.

To be more precise, the deviations with respect to the Paczynski (point-like) curve depend on the brightness profile throughout the source stellar disk. Various profiles have been proposed and discussed in the literature. The first attempts to describe the light intensity variation over the stellar disk, by a linear law [15], were done by Milne and many authors have used that approximation. However, with the advent of more recent stellar atmosphere models it was shown that this simple law was no longer convenient [16–18]. Once it was accepted that the limb-darkening is not a linear phenomenon, alternative laws were proposed: quadratic [19], square root [20], and logarithmic [16] models. Indicating with $I(1)$ the specific light intensity at the center of the stellar disk and with $U$, $a$, $b$, $c$, $d$, $e$, and $f$ the corresponding limb-darkening

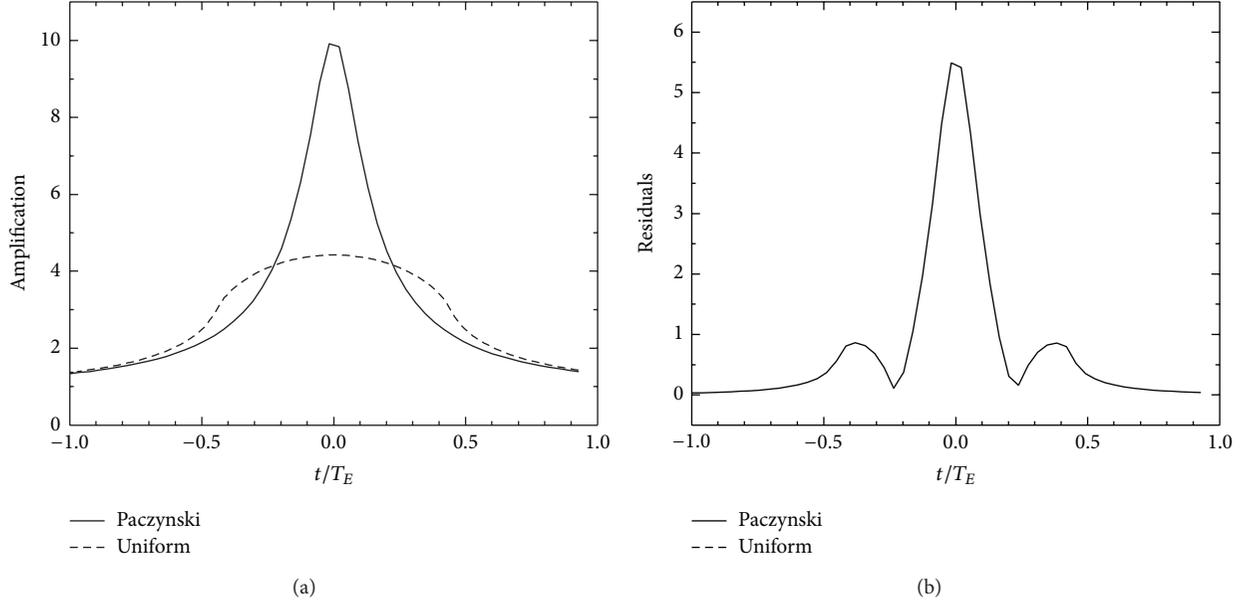

FIGURE 1: (a) Time variation of the amplification for a microlensing event when the source is assumed to be point-like (continuous line) and extended with a uniform brightness profile (dashed line). The event parameters are $u_0 = 0.1$, $\rho_* = 0.46$, and $T_E = 9.2$ hours. The duration of the lens transit over the source disk turns out to be $T \simeq 0.9 T_E = 8.2$ hours, consistent with (1). (b) The residuals between the two curves as a function of time (in units of the Einstein time of the event).

coefficients (LDCs), these models can be expressed by the following laws:

Linear: $I(\mu) = I(1)\left[1 - U(1-\mu)\right]$, (4)

Quadratic: $I(\mu) = I(1)\left[1 - a(1-\mu) - b(1-\mu)^2\right]$, (5)

Square root: $I(\mu)$
$= I(1)\left[1 - c(1-\mu) - d(1-\sqrt{\mu})\right]$, (6)

Logarithmic: $I(\mu)$
$= I(1)\left[1 - e(1-\mu) - f(1 - \mu \ln \mu)\right]$, (7)

with $\mu = \cos\theta$, where $\theta$ is the angle between the line of sight and the emergent intensity. One can therefore calculate the finite source microlensing event magnification. For example, assuming the linear model, one has

$$A_{\rm fs}(u) = \frac{\left(1/\pi\rho_*^2\right)\int_0^{2\pi} d\theta \int_0^{\rho_*} r\left(\left(u^2 + r^2 - 2ur\cos\theta + 2\right)/\sqrt{u^2 + r^2 - 2ur\cos\theta}\sqrt{u^2 + r^2 - 2ur\cos\theta + 4}\right) dr (1 - U\Gamma)}{2\pi \int_0^{\rho_*} r(1 - U\Gamma) dr}, \quad (8)$$

where $\Gamma = 1 - [(\rho_*^2 - r^2)/\rho_*^2]^{1/2}$.

The limb-darkening coefficient $U = [I(\text{centre}) - I(\lim b)]/I(\text{centre})$ gives the light intensity variation from the center of the disk up to its border. For $U = 1$ the star disk is completely limb darkened while for $U = 0$ the star disk is uniformly illuminated. In Figure 2, four light curves corresponding to the same microlensing event with $u_0 = 0.1$ and $\rho_* = 0.46$ are shown: the Paczynski curve corresponding to a point-like source; the light curve for a source with uniform brightness calculated by (3) and two curves calculated by (8) corresponding to $U = 0.2$, $U = 0.7$. For increasing $U$ values, the limb-darkening effects also increase and the obtained light curve approaches closer to the Paczynski profile. As one can easily see, the impact of the exact source brightness profiles, that is, the dependence of the light curve on the LDCs, may be significant in microlensing observations.

In addition to the four limb-darkening profiles above (4), (5), (6), and (7), more recently Claret (2000) has presented a new nonlinear limb-darkening model. Based on the Least-Squares Method (LSM), the obtained brightness profile is able to describe the light intensity distribution in the stellar disk much more accurately than any of the previous models [21]. This model is described by

$$I(\mu) = I(1)\left[1 - a_1\left(1 - \mu^{1/2}\right) - a_2(1 - \mu) - a_3\left(1 - \mu^{3/2}\right) - a_4\left(1 - \mu^2\right)\right] \quad (9)$$

and depends on the four $a_1$, $a_2$, $a_3$, and $a_4$ LDCs.

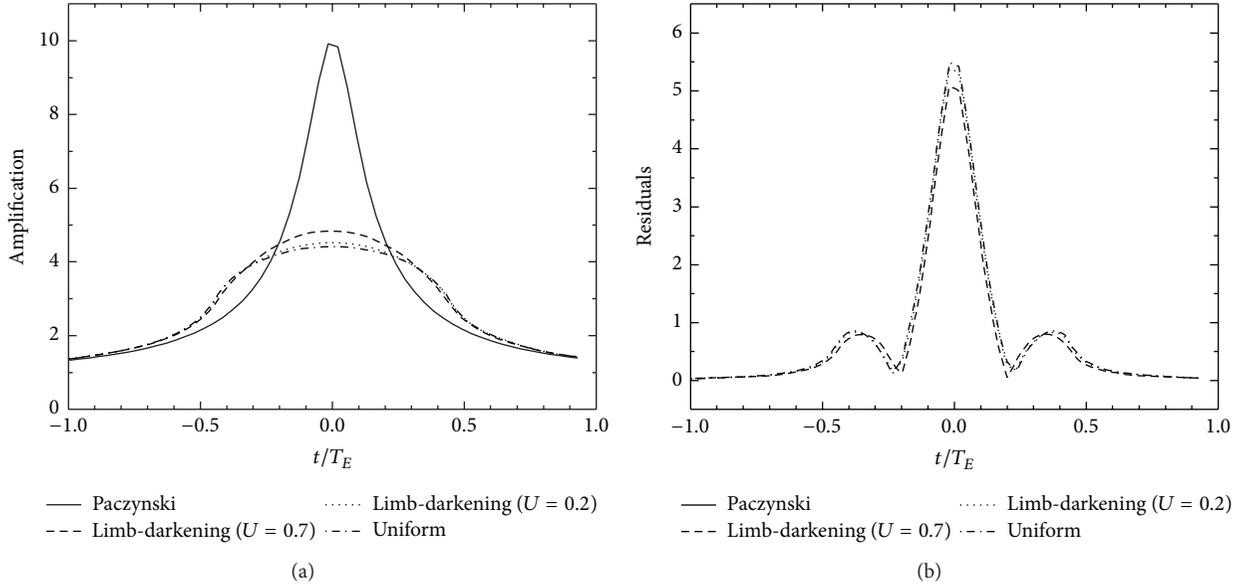

FIGURE 2: (a) Time variation of the amplification for a microlensing event with $u_0 = 0.1$ and $\rho_* = 0.46$ ($u_0 < \rho_*$) in cases: standard (continuous line) uniform source (dot-dashed line) and linear law of LCDs for $U = 0.2$ (dotted line) and $U = 0.7$ (dashed line). Increasing $U$ enhances the limb-darkening effects and thus brings the finite source light curve closer to Paczynski formalism. (b) The residual curves between the Paczynski and the curves with finite source effect are shown. The time is given in units of the Einstein time of the event.

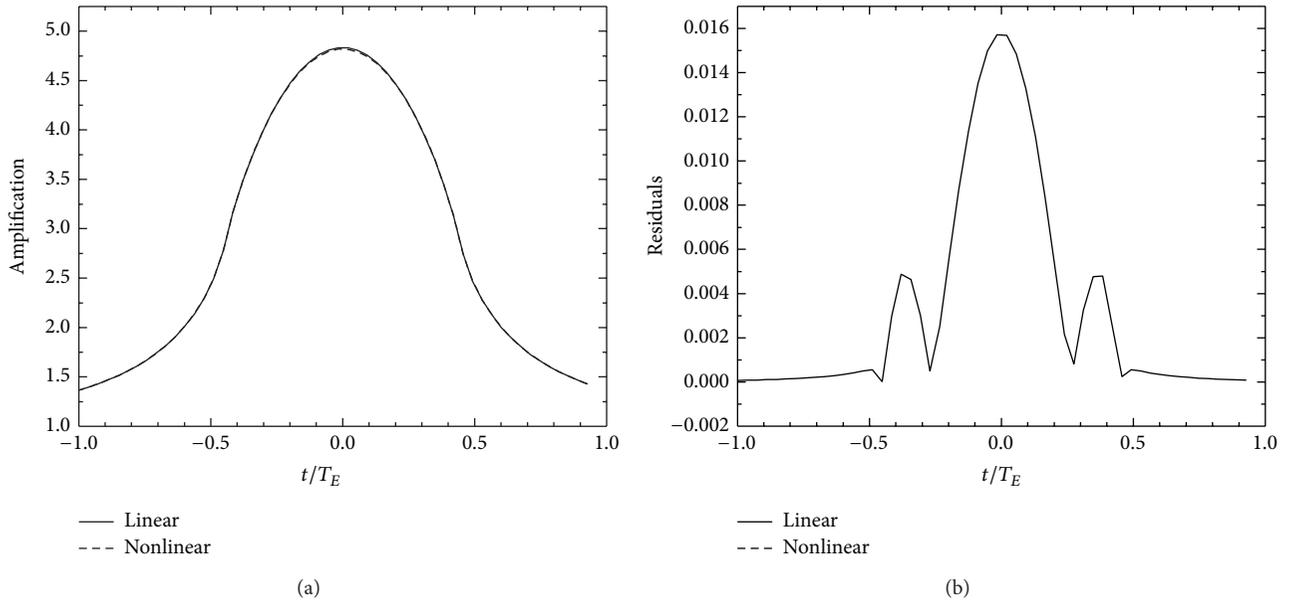

FIGURE 3: (a) Time variation of the amplification for a microlensing event ($u_0 = 0.1$, $\rho_* = 0.46$) with linear LDC ($U = 0.6866$) (continuous line) and with four-parameter nonlinear power law ($a_1 = 0.7181$, $a_2 = -0.5941$, $a_3 = 1.1698$, and $a_4 = -0.4558$) (dashed line). (b) The residual curve between the two curves is shown. The time is given in units of the Einstein time of the event.

A relevant issue is if different microlensing curves induced by different limb-darkening profiles are distinguishable by Euclid telescope. Let us consider a source star at the center of the galaxy with effective temperature $T_{\text{eff}} = 5750$ and $\log g = 4.5$ (i.e., similar to our sun). By Claret, 2000, we find the LDCs in the visible V band filter. As an example, in Figure 3 we plot the obtained microlensing light curves for a linear limb-darkening profile (with $U = 0.6866$) and in the case of a nonlinear profile (with $a_1 = 0.7181$, $a_2 = -0.5941$, $a_3 = 1.1698$, and $a_4 = -0.4558$).

As one can see, the residuals turn out to be about 0.016 at the event peak (in percentage they are up to about 0.32%). Since the Euclid threshold amplification is $A_{\text{th}} = 1.001$ and the photometric error is $\simeq 0.1\%$ [22], these curves can be seen as distinct microlensing profiles by Euclid telescope observations.

## 3. Results

We use a Monte Carlo code to investigate the finite source effects in microlensing events, caused by FFPs in the field of the sky towards the Galactic bulge planned to be observed by the Euclid telescope. Here, we summarize the adopted strategy:

(1) The FFP distance from the Earth $D_L$ is extracted based on the disk and bulge FFP spatial distributions along the Euclid line of sight (the Galactic coordinates are $b = 1.7°$, $l = 1.1°$). The FFP spatial distribution is assumed to follow that of the stars described in [23–25]. We assume an observational cadence of 20 min;

(2) the source distances $D_S$, based on the bulge spatial distributions, result to be in the range from 7 to 10 kpc;

(3) the event impact parameter $u_0$ is assumed to be uniformly distributed in the interval $[0, 6.54]$, not taking into account any selection bias by the experiment [26];

(4) the FFP relative transverse velocity is extracted from a Maxwellian distribution [27, 28];

(5) the lens mass is assumed to follow the mass function defined in [3] with mass function index $\alpha_{PL}$ in the range $[0.9, 1.6]$.

We also use the free code available in the web site http://iac-star.iac.es [29] to generate a synthetic stellar population set of 10000 stars. It allows calculating the stellar population on a star by star basis and computing the luminosity, effective temperature, gravity acceleration $g = GM/R^2$ (that is the star surface gravity), and magnitude of each star. This is performed by a dedicated computer at Instituto de Astrofísica de Canarias (IAC). Based on the *HIT or MISS* algorithm (adapted to two dimensions) [30] we numerically extract 1000 simulated values of $\log g$ and $\log T_{\rm eff}$. Since the four-parameter nonlinear power law is the most versatile limb-darkening law to express the intensity distribution through the source star disk with an excellent accuracy, we use it to obtain the light curve in microlensing events, taking into account finite source effects. Therefore, for each couple ($\log g$, $\log T_{\rm eff}$) the four LDCs $a_1, a_2, a_3$, and $a_4$ are obtained from the table available at the web site http://vizier.u-strasbg.fr/viz-bin/VizieR-3?-source=J/A%2bA/363/1081/phoenix [21].

Assuming the standard mass-luminosity relation $L/L_\odot = (M/M_\odot)^\alpha$ (with $\alpha \simeq 3.5$) and considering that the main sequence star luminosity is given with a sufficiently good approximation by the Stefan Boltzmann law $L = 4\pi R^2 \sigma T_{\rm eff}^4$, we find the relation

$$\frac{R}{R_\odot} = \left\{ \left(\frac{T_{\rm eff}}{T_{\rm eff\odot}}\right)^4 \left(\frac{g_\odot}{g}\right)^\alpha \right\}^{1/(2\alpha-2)}, \quad (10)$$

where $T_{\rm eff\odot} = 5778$ K and $g_\odot = 27381.2$ cm/s$^2$ ($\log g_\odot = 4.4374$).

Once all the parameters are fixed, we can calculate for each event the Paczynski curve, the more realistic light curve that takes into account the finite source effects and the residuals between them. Considering the performances of the Euclid telescope, we sampled each light curve with exposures taken every 20 min, which means that an event must have a duration longer than 2.67 hours to be detectable. We assume that a microlensing event can be detected if in its light curve there are at least eight consecutive points with the amplification larger than the threshold amplification $A_{\rm th} = 1.001$. Finite source effects can be detected on a light curve when it contains at least eight points with $\text{Re } s > 0.001$ around the event peak (see [22]). We retain all synthetic events with residuals fulfilling the above-mentioned conditions. Therefore, the efficiency for finite source effect detection is given by the ratio between the number of these events and the total number of detectable events.

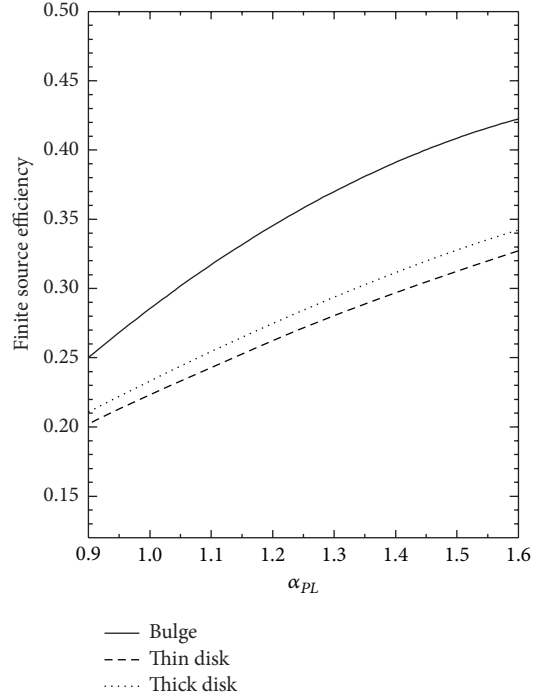

FIGURE 4: The finite source efficiency for microlensing events by FFPs as a function of $\alpha_{PL}$ for the three different distributions of FFPs: bulge (continuous line), thin disk (dashed line), and thick disk (dotted line) considered separately.

In Figure 4 we show our results for the finite source efficiency in microlensing events caused by FFPs and expected to be detectable by Euclid teleskope. We have considered separately the lens distribution as bulge, thin disk, and thick disk objects and define the finite source efficiency with respect to the value of $\alpha_{PL}$ (ranging from 0.9 to 1.6). The efficiency for finite source effect detection is larger (in the range 25% to 42%) for bulge FFPs and slightly lower for thin and thick disk events (approximately in the range 20% to 35%) but always tends to increase with increasing $\alpha_{PL}$ values.

So, one can say that in approximately 30% of the microlensing events caused by FFPs it will be possible to detect the finite source effects in the light curves by using Euclid observations. This certainly constitutes a nonnegligible number of the expected microlensing events due to FFPs.

TABLE 1: For different populations of FFPs in the Galactic bulge and thin and thick disks and assuming the mass function index $\alpha_{PL} = 1.3$, we give the percentage of expected events in each indicated bin for the magnification peak value.

| $A_{\max} \rightarrow$ | [1.001, 5[ | [5, 10[ | [10, 15[ | [15, 20[ | [20, 25[ | [25, 30[ | >30 |
|---|---|---|---|---|---|---|---|
| Bulge | 12.25% | 1.15% | 0.20% | 0.30% | 0.05% | 0.25% | 0.05% |
| Thin disk | 7.10% | 1.45% | 0.40% | 0.25% | 0.20% | 0.05% | 0.55% |
| Thick disk | 5.85% | 0.80% | 0.15% | 0.30% | 0.20% | 0.15% | 0.35% |

The efficiency obtained with the procedure described previously coincides with that obtained by considering the fraction of transit events for which the projected lens on the source plane crosses the surface of the source star, those with $\rho_* \geq u_0$ in (1) to the simulated ones. For example, for $\alpha_{PL} = 1.3$, these events are about 39% of the simulated ones for bulge FFPs and slightly less than 30% for FFPs in the thin and thick disks of the Milky Way.

In Table 1 we show, for FFPs with $\alpha_{PL} = 1.3$ in the bulge and thin and thick disks, the fraction of the events (in percentage) with detectable finite source effects by the Euclid telescope observations with peak magnification in each defined bin range.

It is clear from Table 1 that the obtained percentage of events with detectable finite source signatures in the light curves tends to decrease as the magnification peak value of the light curve increases. The efficiency to detect finite source effects in highly magnified events is less than 1%. However, taking into account that Euclid observations will allow detecting $\sim 10^3$ microlensing events per month and $\sim 100$ events per month due to FFPs [22, 31], there will be a nonnegligible chance to detect finite source effects even in highly amplified events.

A significant point to consider is how the finite source efficiency depends on $x = D_L/D_S$. To answer this question we simulated 1000 microlensing events for each defined bin range [0, 0.1]–[0.9, 1] of $x$, with $\alpha_{PL} = 1.3$, and calculated the finite source efficiency as defined above. Our results, for FFPs in the bulge and in the thin and thick Galactic disks, are shown in Figure 5. We can see the finite source efficiency increases with the value of $x$, so it is more likely to detect finite source effects for FFPs in the Galactic bulge. For FFPs in the thin and thick disks we also note that the finite source efficiency slightly diminishes in the [0.4, 0.5] bin. This is not due to a numerical fluctuation but is related to the fact that $R_E$ gets its maximum value at $x = 0.5$ (see (1) at [22]). We also would like to emphasize that there will even be with Euclid observations a chance to detect finite source effects in FFP events relatively close to the observer, that is, with $x \leq 0.1$. These events, although rare, are particularly interesting since they might allow a direct observation of the FFP that microlensed a star, as what already happened for the event MACHO-LMC-5 [32].

## 4. Conclusions

In this paper, we have considered the deviations due to finite source effects in the microlensing light curves caused by FFPs toward the planned Euclid field of view towards the Galactic bulge. We emphasize here that if the planned Euclid program for microlensing observation towards the Galactic bulge will be effectively performed, the unique opportunity to detect finite source effect in very short microlensing events will become real.

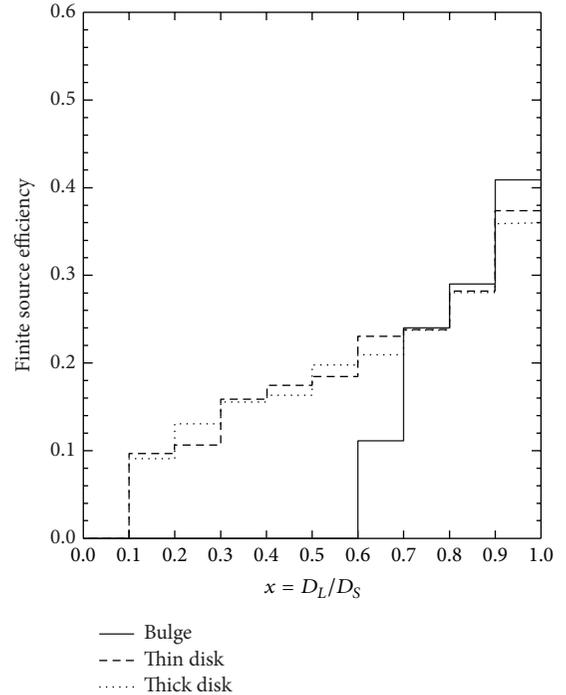

FIGURE 5: The finite source efficiency for microlensing events by FFPs as a function of $x$ for the three different distributions of FFPs: bulge (continuous line), thin disk (dashed line), and thick disk (dotted line). The plot is for $\alpha_{PL} = 1.3$.

In our calculations it is assumed that the source stars appear with circular shape on the sky and that their limb-darkening profile is given by the four-parameter nonlinear power law.

We also note that, as discussed in Section 3, Euclid telescope data may allow distinguishing among different limb-darkening profiles.

Using Monte Carlo numerical simulations, we investigate the efficiency to detect finite source effects in microlensing light curves caused by FFPs. In Figure 4, we have shown that this effect is potentially interesting since finite source effects turn out to be detectable in about 1/3 of all observable events if the FFP mass index is $\alpha_{PL} = 1.3$.

Moreover, if $\alpha_{PL}$ gets larger values, the efficiency of detecting finite source effects further increases. Since Euclid teleskope will detect a considerably large number (about 100 per month) of microlensing events produced by FFPs towards the Galactic bulge [22, 31], the number of events with detectable finite source effects may become quite high and will allow constraining the FFP mass and distance distribution, which is fundamental information necessary to investigate FFP populations throughout the Milky Way. This, in turn, is an important issue in order to establish their origin. Moreover, high-magnification microlensing events caused by FFPs, although rare (about 1% of the FFP events), will be likely discovered by Euclid teleskope, offering the unique possibility to investigate the Galactic bulge star atmospheres and to define the source limb-darkening profile.

## Conflict of Interests

The authors declare that there is no conflict of interests regarding the publication of this paper.

## Acknowledgment

Francesco De Paolis and Achille A. Nucita acknowledge the support by the INFN project TAsP.

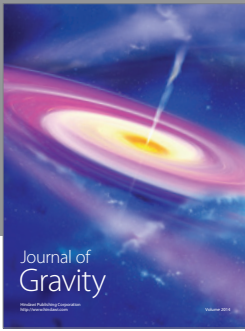 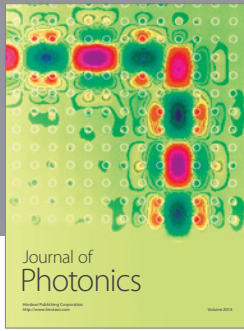 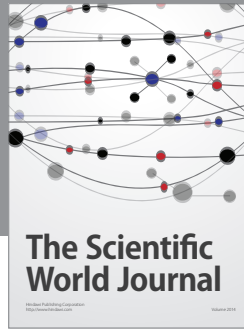 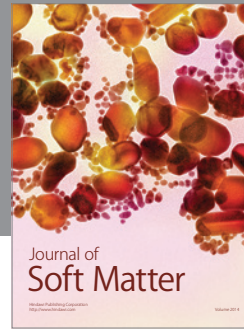 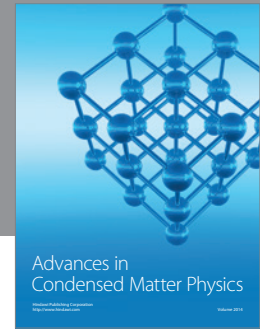 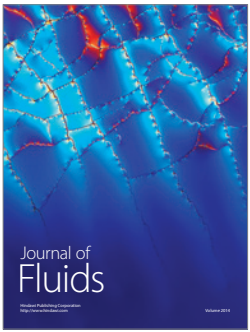 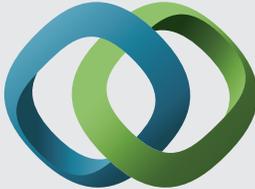 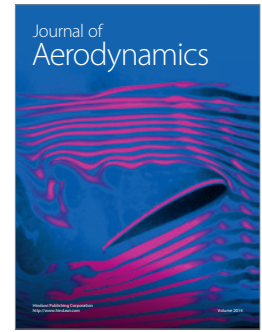 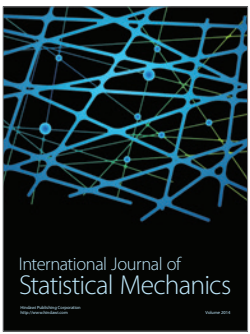 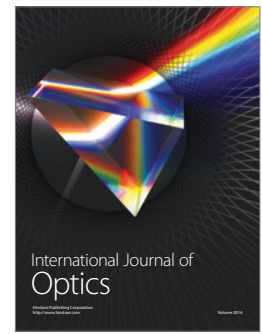 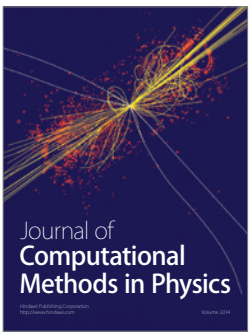 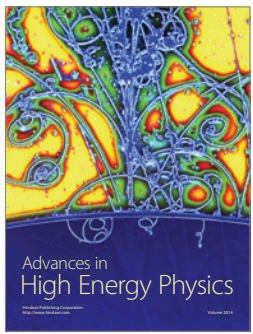 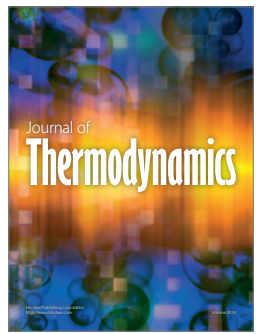 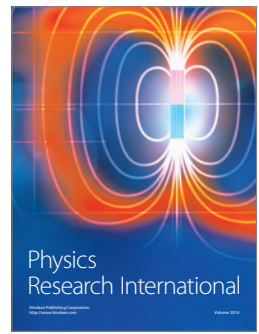 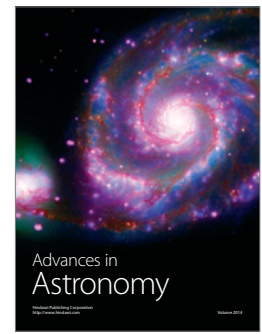 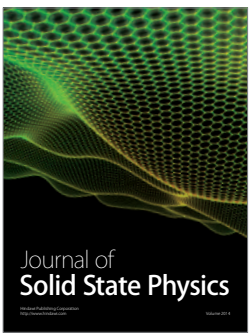 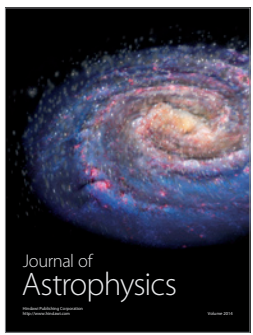 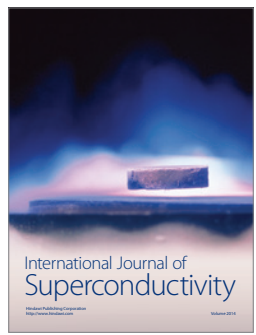 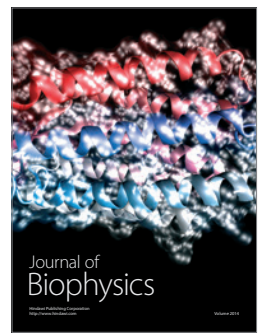 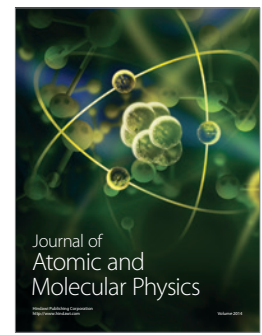